\documentclass{cernyrep} 
\usepackage{texnames}
\usepackage[T1]{fontenc}
\usepackage{wrapfig}
\begin{document}
\title{LHC Results - Highlights}
 
\author{Gigi Rolandi}

\institute{CERN, Geneva, Switzerland and Scuola Normale Superiore, Pisa, Italy.}

\maketitle 

\begin{abstract}
 The LHC has delivered already 10 ${\rm fb}^{-1}$ of proton proton collisions at a centre-of-mass energy of 7-8 TeV.
 With this data set, ATLAS and CMS have discovered a new boson at a mass of about 125 GeV and have searched for new physics at the TeV scale. 
\end{abstract}

 \section{Introduction}
 
The LHC~\cite{Evans:2008zzb} performs well above expectations with a peak luminosity of $4\, 10^{33}\, {\rm cm}^{-2} {\rm s}^{-1}$ from collisions of two 3.5 TeV proton beams in 2011 and two 4 TeV beams in 2012 . In 2011, the LHC delivered about 6 ${\rm fb}^{-1}$ to ATLAS~\cite{Aad:2008zzm} and CMS~\cite{ref:CMS}.  With this large integrated luminosity, it is possible to search effectively for the Standard Model (SM) Higgs boson and to probe the existence of new physics at the TeV scale.  This luminosity has been reached with an inter bunch spacing of 50 ns and more than $1.2\;10^{11}$ protons per bunch. At these currents the average number of inelastic interactions per bunch crossing (pile-up) is about 15, hence posing new challenges to trigger, event reconstruction, and quality of reconstructed physics objects.  

New algorithms have been designed to mitigate the effect of the pileup. In spite of the difficult experimental conditions, ATLAS and CMS have been able to calibrate quickly their data and to deliver new physics results shortly after the start of data taking. They are producing physics papers at a rate of about 100 papers per year per experiment, probing the Standard Model and searching for new physics. The main physics messages of the analyses of the data collected up to now are:
\begin {itemize}
\item the Standard Model  is still in excellent shape;
\item a new boson~\cite{Aad:2012gk, Chatrchyan:2012gu}  has been found with properties compatible with those predicted for the SM Higgs Boson;
\item  no sign of new physics has been found yet.
\end{itemize}

In this lecture, I will concentrate on two topics : search for Supersymmetry and search for the Higgs boson. 

\section{Supersymmetry}
Supersymmetry (SUSY)~\cite{ref:SUSY-1, ref:SUSY0, ref:SUSY1,
ref:SUSY2, ref:SUSY3, ref:SUSY4, ref:hierarchy1, ref:hierarchy2}
is a well motivated extension of the SM.
It introduces a large number of new particles with the 
same quantum numbers as their SM partners, but differing by half a unit of spin. 

With R-parity conservation~\cite{Farrar:1978xj}, the supersymmetric
particles, such as squarks and gluinos, are produced in pairs and
decay to the lightest, stable supersymmetric particle (LSP). If the
LSP is neutral and weakly interacting, a typical signature is a final
state of multi-jets and possibly leptons accompanied by large Missing 
Transverse Energy (MET). 

\begin{figure}
\begin{center}
\includegraphics[width=0.4\textwidth]{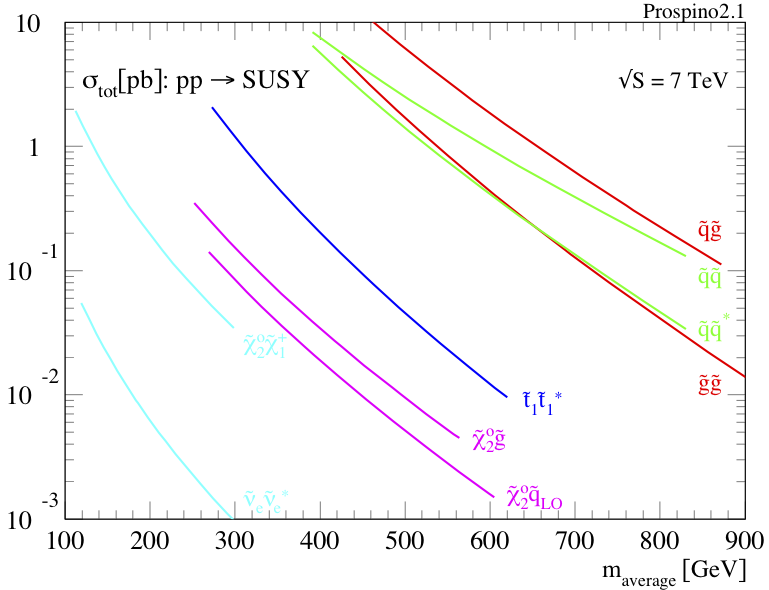}
\end{center}
\caption{Cross sections for producing SUSY particles in pp collisions at $\sqrt{s}=$ 7 TeV computed with PROSPINO\cite{Beenakker:1996ch,Beenakker:1997ut,Beenakker:1999xh}}
\label{fig:prospino}
\end{figure}

The cross sections for producing SUSY particles are shown in Fig.~\ref{fig:prospino} as a function of the mass of the particles. In quark and gluons collisions it is easy to produce coloured objects like gluinos and squarks, which decay typically to jets and MET, while the cross sections for Electroweak productions are smaller and the mass reach substantially reduced. These "ewkinos" decays typically produce many leptons and MET.


Both ATLAS~\cite{Aad201267,ATLASJetMET,ATLAS2011,Aad:2012hm,Aad:2012rz} and 
CMS~\cite{RA2paper35pb-1,RAZORpaper35pb-1,AlphaTPaper35pb-1,AlphaTPaper1fb-1,Chatrchyan:2012jx} presented many hadronic SUSY searches on the 7 TeV data based directly or indirectly on MET.  These searches can be interpreted in many ways.  Simplified versions of SUSY, with a drastic reduction of the more than 100 parameter space like CMSSM~\cite{ref:CMSSM} or mSugra~\cite{ref:MSUGRA}, are excluded for gluinos and squarks below about 1 TeV and are now cornered. The searches can also be interpreted in terms of simplified models~\cite{Alves:2011wf} where a single decay chain is considered with the assumption that the branching fractions along this chain are 100\%.

\begin{figure}[h!]
\begin{tabular} {cc}
\includegraphics[width=0.48\textwidth]{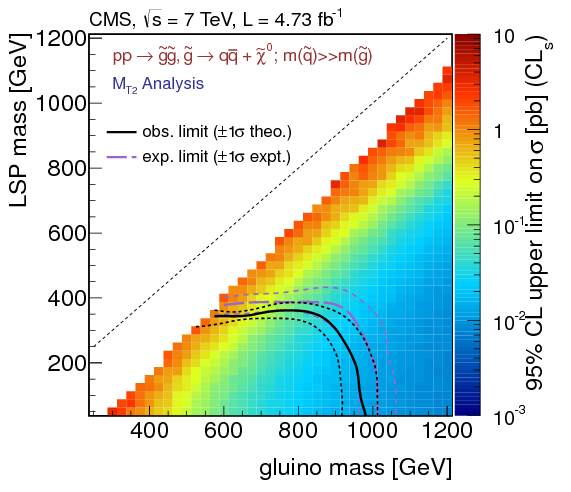}
&
\includegraphics[width=0.42\textwidth]{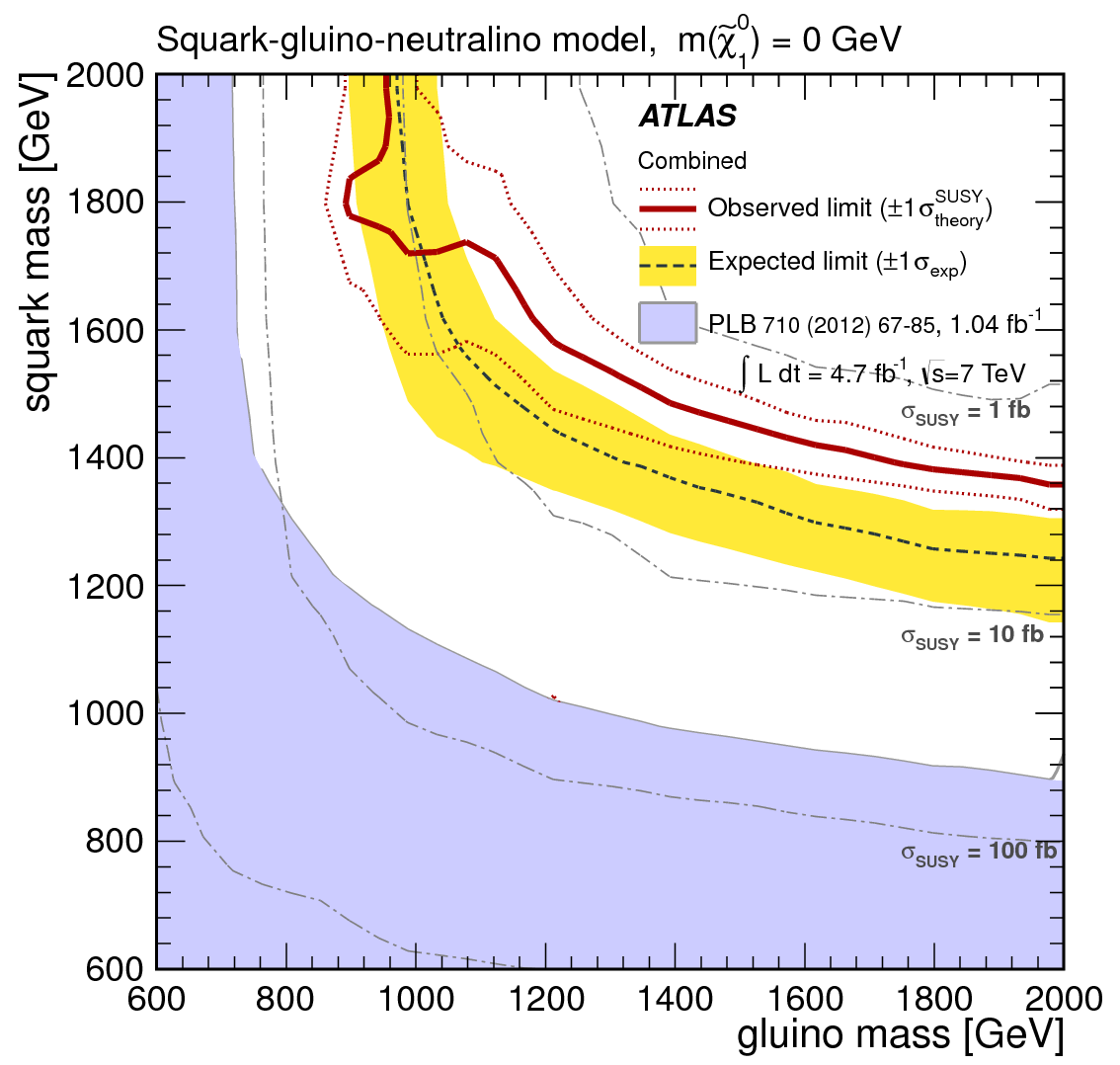}
\end{tabular}
\caption{Upper limit~\cite{Chatrchyan:2012jx} on the cross section for gluino pair production when squarks are much heavier than gluinos and the gluinos decay to two light quarks and a neutralino (left). Upper limit~\cite{Aad:2012rz} on squark-gluino associated production with the gluino decaying into a quark pair and a neutralino and the squark decaying into quark neutralino. The neutralino here is assumed to be massless(right).}
\label{fig:sms}
\end{figure}

Figure~\ref{fig:sms} shows two examples. The models assumed here are i) gluino pair production when squarks are much heavier than gluinos and the gluinos decay to two light quarks and a neutralino (left) and ii) squark gluino associated production with gluino decaying into a quark pair and a neutralino and the squark decaying into quark neutralino. The neutralino here is assumed to be massless (right).

\begin{figure}[h!]
\begin{tabular} {cccc}
\includegraphics[width=0.22\textwidth]{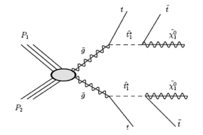}
&
\includegraphics[width=0.25\textwidth]{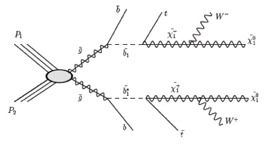}
&
\includegraphics[width=0.22\textwidth]{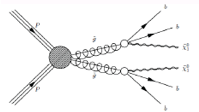}
&
\includegraphics[width=0.22\textwidth]{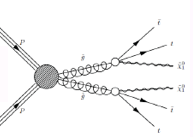}
\\
(a)&(b)&(c)&(d)
\end{tabular}
\caption{Simplified models used for interpretations of searches for production of third generation squarks~\cite{Chatrchyan:2012sa,Aad:2012pq}}
\label{fig:sms1}
\end{figure}

While generic SUSY production at the scale of about 1 TeV is not compatible with data, there is still room for natural models~\cite{Barbieri:1987fn,Papucci:2011wy} where gluinos and third-generation squarks are below 1 TeV. 
Both ATLAS and CMS have done specific searches for the third-generation squarks. Examples are the search for same-sign lepton pairs, b jets and MET by CMS~\cite{Chatrchyan:2012sa} - motivated by final states including four top  or two top and two W  as shown in Fig.~\ref{fig:sms1} (a) and (b) - and the search for three b-jets and MET by ATLAS~\cite{Aad:2012pq} addressing models shown in Fig~\ref{fig:sms1} (c) and (d). Broadly speaking, these searches exclude gluinos of 900 GeV for third generation squarks lighter than 300-400 GeV.

\begin{wrapfigure}{r}{0.5\textwidth}
\vspace{-20pt}
\begin{center}
\includegraphics[width=0.4\textwidth]{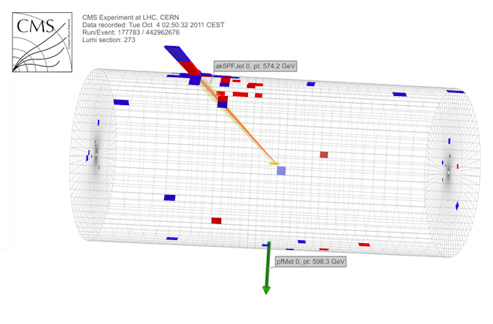}
\end{center}
\vspace{-20pt}
\caption{Monojet event recorded by CMS}
\label{fig:cms_wimp}
\end{wrapfigure}

Both ATLAS and CMS have also performed a model-independent search for Weakly Interacting Massive Particles (WIMP) production~\cite{Aad:2011xw,Chatrchyan:2012me} triggering on events with a monojet and MET. Here the jet is produced by initial state radiation of one of the interacting partons and the two WIMPs escape undetected leading to spectacular events like the one shown in Fig.~\ref{fig:cms_wimp}: the detector is empty with the exception of a single high energy jet. In the SM these events are produced by a high transverse momentum  Z decaying into a neutrino pair. This background can be effectively measured from similar events where the Z decays into a muon pair. 

The measured cross section is compatible with the SM background and limits can be set on generic WIMP production.   For dark matter models, the observed limit on the cross section depends on the mass of the dark matter particle and the nature of its interaction with the SM particles. The limits on the effective contact interaction scale as a function of  the wimp mass can be translated into a limit on the dark matter-nucleon scattering cross section\cite{bib:RoniHarnik}. These limits can be compared with the constraints from direct and indirect detection experiments. The LHC limits~\cite{Chatrchyan:2012me} are competitive with those from direct WIMP search for the spin-dependent interaction for WIMP mass below few hundreds GeV, and also for the spin-independent interaction for WIMP mass below 10 GeV .

\section{Search for the Higgs boson}
Three weeks after I gave this lecture in Anjoux, ATLAS~\cite{Aad:2012gk} and CMS~\cite{Chatrchyan:2012gu} announced the observation of a new boson compatible with the SM Higgs boson in a CERN seminar. In this section, with the agreement of the organizers of the school, I describe the observation of the boson.

The search for the Higgs boson (H) and the justification of the electroweak symmetry breaking~\cite{Englert:1964et,Higgs:1964pj} was one of the main reasons for the construction of the Large Hadron Collider. This search was therefore a high priority analysis for ATLAS and CMS. In the SM the cross section for H production in proton proton collision at 7 TeV is about  17 pb~\cite{LHCHiggsCrossSectionWorkingGroup:2011ti} for  ${\rm m}_{\rm H} \simeq 125$ GeV and 30\% higher at 8 TeV. In about 10 fb$^{-1}$ of integrated luminosity, shared about equally between the two energies, some 200,000 Higgs bosons are produced in each experiment. It is very difficult however to separate this signal from the very large background of SM processes, especially the hadronic final states, and specific searches with leptons or photons in the final state are performed. 
Table~\ref{tab:TH1} lists the search channels together with the branching ratio expected for a SM Higgs boson.

\begin{table}[b]
\begin{center}
\caption{Channels used in the search for the Higgs boson and the branching ratios expected in the SM for ${\rm m}_{\rm H} =125$ GeV. Here $\ell$ and $\ell'$ indicate an electron or a muon.}
\label{tab:TH1}
\begin{tabular}[h!]{lc}\hline\hline
channel & Branching Fraction \\\hline
${\rm H }\rightarrow {\rm ZZ} \rightarrow \ell \ell \; \ell' \ell' $ & $1.2\; 10^{-4}$\\
${\rm H } \rightarrow \gamma \gamma $ & $ 2.3\; 10^{-3}$\\
${\rm H } \rightarrow {\rm WW} \rightarrow \ell \nu \; \ell' \nu' $ & $1.0\; 10^{-2}$\\
${\rm H } \rightarrow \tau \tau $ & $6.0\; 10^{-2}$\\
${\rm H } \rightarrow {\rm bb} $ & $5.8\; 10^{-1}$\\
\hline\hline
\end{tabular}
\end{center}
\end{table}

A value of ${\rm m}_{\rm H}\simeq 125$ GeV is smaller than the sum of the masses of the vector boson (V) pairs, in the decays ${\rm H} \rightarrow {\rm VV}$ one or both V are off mass-shell. In  ${\rm H}\rightarrow 4\ell$ and ${\rm H} \rightarrow \gamma \gamma $ the H mass is reconstructed with high resolution (1-2\%) with the precise measurements of the momenta of leptons and photons. Because the width of the SM Higgs boson for ${\rm m}_{\rm H} \simeq 125$ GeV  is a few MeV, one expects to see a narrow peak dominated by the instrumental resolution. The other channels have worse mass resolution because  of the missing neutrinos in the W and $\tau$ decays or because the b-jets are reconstructed with some 10\% resolution. The search in the bb final state is performed in the associate production ${\rm VH}\rightarrow {\rm Vbb}$ where the V decays into leptons that provide the trigger and reduce the overwhelming hadronic background.

\begin{wrapfigure}{r}{0.5\textwidth}
\vspace{-20pt}
\begin{center}
\includegraphics[width=0.4\textwidth]{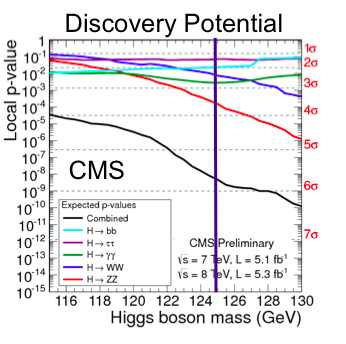}
\end{center}
\vspace{-20pt}
\caption{Discovery potential of CMS: p-value as function of the Higgs boson mass }
\label{fig:expected}
\vspace{-40pt}
\end{wrapfigure}

Figure~\ref{fig:expected} displays the expected discovery potential for a SM Higgs boson in the CMS experiment as a function of ${\rm m}_{\rm H}$ (ATLAS have similar figures).  The probability that the background can produce a fluctuation greater than the potential excess produced  in data by a SM Higgs boson (the so called local p-value)  is estimated less than $10^{-8}$ i.e. more than five standard deviations. The most sensitive channels are those where a narrow peak can be observed: ${\rm H}\rightarrow 4\ell$ and ${\rm H}\rightarrow \gamma \gamma$. The WW channel is also quite sensitive.

\subsection{${\rm H} \rightarrow {\rm WW} \rightarrow \ell \nu \; \ell' \nu' $}

\begin{figure}[h!]
\begin{tabular} {cc}
\includegraphics[width=0.48\textwidth]{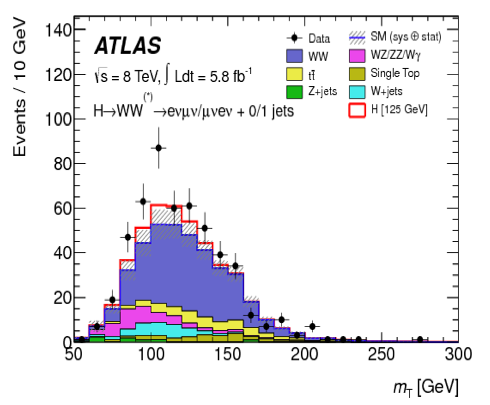}
&
\includegraphics[width=0.48\textwidth]{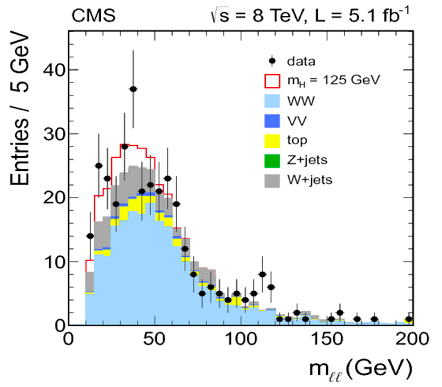}

\\
(a)&(b)
\end{tabular}
\caption{ ${\rm H} \rightarrow {\rm WW}$ analysis. a) ATLAS, Distribution of the transverse mass, in the 0-jet and
1-jet analyses with opposite flavour, for events satisfying
all selection criteria. The expected signal for ${\rm m}_{\rm H}$ = 125 GeV is
shown added to the background prediction.  The hashed
area indicates the total uncertainty on the background prediction. b) CMS, Distribution of dilepton invariant mass for the zero-jet opposite flavour category  at 8 TeV after the full selection, except for the selection on $m_{\ell \ell}$ itself. The signal expected from a Higgs boson with a mass ${\rm m}_{\rm H}$ = 125 GeV is shown added to the
background.}
\label{fig:ww}
\end{figure}

The decay mode  ${\rm H} \rightarrow {\rm WW} \rightarrow \ell \nu \; \ell' \nu' $ is the main search channel for  a SM Higgs boson with mass above the WW  threshold of 160 GeV. With good experimental control of the MET and very tight lepton identification, it is possible to reject large part of the reducible background and extend the sensitivity down to ${\rm m}_{\rm H} \simeq 120$  GeV.
The signature is two isolated opposite-sign charged leptons and large MET caused by the two undetected neutrinos. The most sensitive channel is when the two leptons have opposite flavour and there are no extra jets in the event. Here the main background is the irreducible non resonant WW production and the reducible W+jet production when the jet fakes a lepton. The other channels with same flavour leptons or with associated jets have larger backgrounds from Drell-Yan and top quark decay respectively and contribute less than 20\% to the sensitivity. The yields  of the largest backgrounds are estimated from control regions. One important variable to separate the signal from the irreducible background is the angle between the two leptons. Due to spin correlations, this variable is small for W pairs from the spin-0 H decay and large for the WW non resonant production~\cite{Dittmar:1996ss}. In ATLAS the transverse mass of the MET vector and the di-lepton system - shown in Fig.~\ref{fig:ww} a - is used to test for the presence of a signal for all jet multiplicities. In CMS, the signal is separated from the background with kinematical and topological requirements optimized for each mass hipothesis; one of the most sensitive variables is the dilepton invariant mass shown in  Fig.~\ref{fig:ww} b.

In the most sensitive channel, leptons with opposite flavour and zero jets, CMS observe 158 events,  estimate a background of $124 \pm 12$ events and expect $24\pm 5 $ events for a SM Higgs boson of 125 GeV. Similarly ATLAS observe 185 events,  estimate a background of $142 \pm 16$ events and expect $20\pm 4$ events for a SM Higgs boson of 125 GeV.

\subsection{${\rm H} \rightarrow {\rm ZZ} \rightarrow \ell \ell \; \ell' \ell' $ }

The decay mode ${\rm H} \rightarrow {\rm ZZ} \rightarrow \ell \ell \; \ell' \ell' $ , the so called golden channel, is characterized by a small signal yield over a flat irreducible background of direct ZZ production. Because the signal yield is small, it is important to maximize the efficiency lowering as much as possible the threshold on the lepton transverse momenta: at ${\rm m}_{\rm H}= 125$ GeV the average transverse momentum of the softest lepton is about 7 GeV. Also the lepton identification is relaxed in order to maximize the efficiency. The reducible background is evaluated with control regions and is small, in spite of the relaxed identification criteria, thanks to the presence of four leptons in the final state.

In the ATLAS analysis the thresholds for muons and electrons are 6 and 7 GeV. In CMS, they are 5 and 7 GeV. The events are selected pairing opposite charge and same flavour leptons and with requirements on the invariant masses of the pairs. A recovery of the final state radiation photons is also used in CMS . The expected background yield of the irreducible background is estimated using the MC simulation normalized to the theoretical cross section for ZZ production. The identification efficiency, the energy scale and the energy resolution are measured using large samples of Z , Y and J/$\psi$ decaying into two leptons.

\begin{figure}[h!]
\begin{tabular} {cc}
\includegraphics[width=0.48\textwidth]{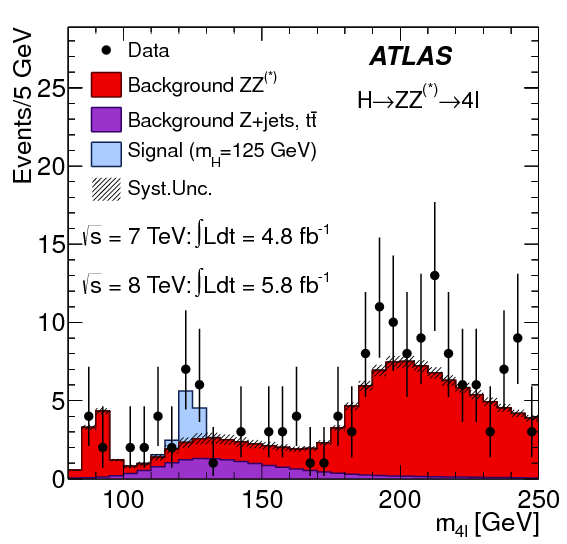}
&
\includegraphics[width=0.48\textwidth]{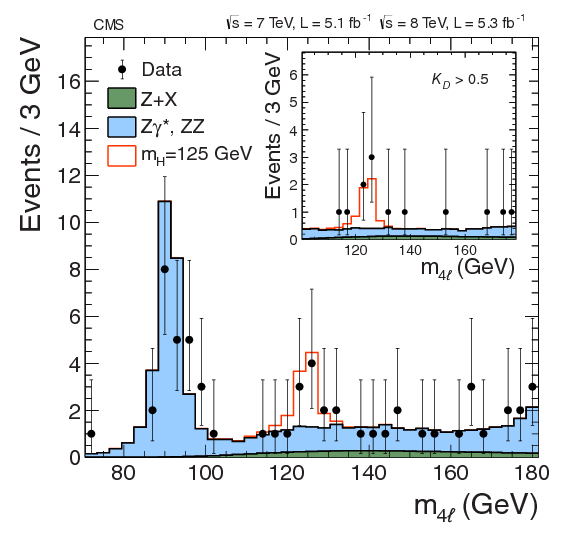}

\\
(a)&(b)
\end{tabular}
\caption{ ${\rm H} \rightarrow {\rm ZZ} \rightarrow \ell \ell \; \ell' \ell' $. Distribution of the four-lepton invariant mass for the selected candidates, compared to the background expectation. The signal expectation for a SM Higgs boson with ${\rm m}_{\rm H}=125$ GeV is also shown. (a) ATLAS,  (b) CMS;  the inset shows the distribution after selection of events with KD > 0.5, as described later in the text.}
\label{fig:4leptons}
\end{figure}

 Figure~\ref{fig:4leptons} shows the invariant mass distribution of the selected events compared to the estimated background. The peak at about 90 GeV is the decay of the Z into four leptons, where a lepton pair is radiated by one of the leptons originating from the Z decay. In this process the distribution of the lowest momentum lepton is softer than in the Higgs boson decays. The Z peak has a different yield in the two experiments because of the larger efficiency in CMS for low momentum leptons. The reducible background in CMS is indeed much smaller than the reducible one, in ATLAS they are comparable for ${\rm m}_{\rm H}=125$ GeV.
 
\begin{figure}[h!]
\begin{tabular} {cc}
\includegraphics[width=0.48\textwidth]{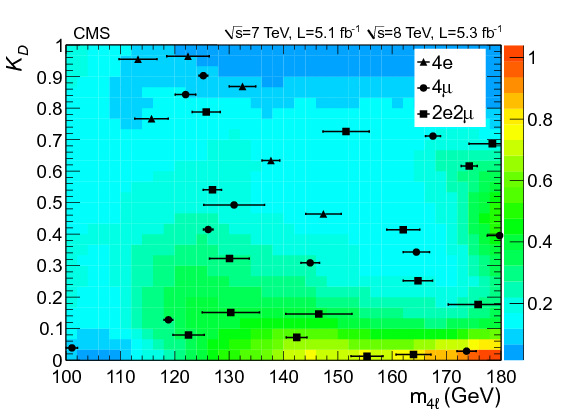}
&
\includegraphics[width=0.48\textwidth]{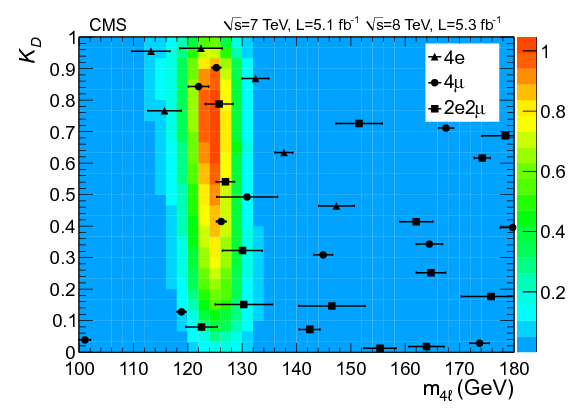}

\end{tabular}
\caption{ ${\rm H}\rightarrow {\rm ZZ} \rightarrow \ell \ell \; \ell' \ell' $. CMS. The distribution of events
selected in the four lepton channels for the kinematic discriminant, KD, versus four lepton mass.
Events are marked by symbols.
The horizontal error bars indicate the estimated mass resolution.
In the left plot the colours show the expected background;
in the right plot the colours show the event density expected from a SM Higgs boson of 125 GeV
(both in arbitrary units).}
\label{fig:mela}
\end{figure}

In the four-lepton mass region  121.5-130.5 GeV CMS observe 9 events,  estimate a background of $3.8 \pm 0.5$ events and expect $7.5\pm 0.9 $ events for a SM Higgs boson of 125 GeV. In the four-lepton mass region  120-130 GeV ATLAS observe 13 events,  estimate a background of $4.8 \pm 0.2$ events and expect $5.3\pm 0.5$ events for a SM Higgs boson of 125 GeV. 

The scalar nature of the Higgs boson provides important discriminating power between the signal and the irreducible background. 
The kinematics of the ${\rm ZZ} \rightarrow \ell \ell \; \ell' \ell'$ process is fully  described by five angles and the invariant masses of the two lepton pairs~\cite{Cabibbo:1965zz,Gao:2010qx,DeRujula:2010ys} for a fixed invariant mass of the four-lepton system. In CMS, a kinematic discriminant  (KD) is constructed based on the probability ratio of the signal and background hypotheses as described in Ref.~\cite{Chatrchyan:2012sn}. Figure~\ref{fig:mela} shows the the distribution of four leptons invariant mass versus KD for the selected events.
The discriminant KD takes large values for signal like events and small values for background like events.

 A clustering of events is observed with a high value of the kinematic discriminant at ${\rm m}_{\rm H} \simeq 125$ GeV where the background expectation is low,  corresponding to the excess seen in the one-dimensional mass distribution.

\subsection{${\rm H} \rightarrow \gamma \gamma $}
The decay mode ${\rm H} \rightarrow \gamma \gamma $ is characterized by  a narrow peak in the diphoton invariant mass distribution above a large irreducible background from QCD production of two photons and a reducible background where one - and in few cases two - reconstructed photons originate from misidentification of jets. The resolution in the invariant mass of the two photons varies on event by event basis, depending on the properties of the reconstructed photons and of the overall event properties (e.g. number of reconstructed vertices). In order to improve the sensitivity of the analysis the events are categorized in exclusive sets with varying signal purity. The events with two reconstructed jets belong to separate categories to exploit the better signal to background ratio in the H production via Vector Boson Fusion (VBF) where two jets  originating from the two scattered quarks are expected with large difference in rapidity.

\begin{figure}[h!]
\begin{tabular} {cc}
\includegraphics[width=0.48\textwidth]{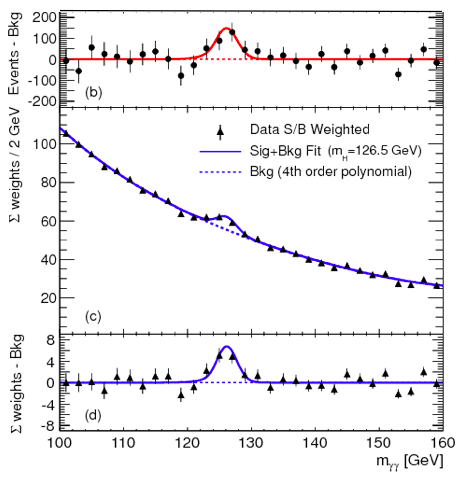}
&
\includegraphics[width=0.48\textwidth]{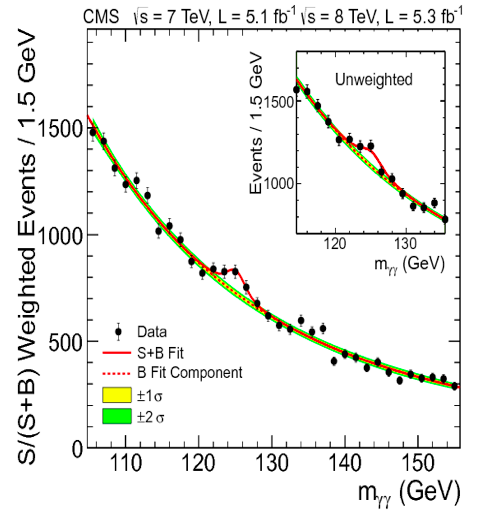}
\end{tabular}
\caption{ ${\rm H} \rightarrow \gamma \gamma $. The distributions of the invariant mass of diphoton candidates
after all selections are shown together with the result of a fit to the data of the sum of a signal component and a background component. (left) ATLAS.  The panel in the middle (c) shows the weighted sample with the weights are explained in the text. The  signal is fixed to ${\rm m}_{\rm H} = 126.5$ GeV  and the background is described by a fourth-order Bernstein polynomial . The residuals of the weighted data with  respect to the fitted background component are displayed  in (d). The panel on the top (b) shows the residuals of the un-weighted data with respect to their fitted background. (right) CMS. The panel shows the weighted sample where the background is fitted with a fifth order polynomial . The coloured bands represent the $\pm1$ and $\pm2$ deviation uncertainties on the background estimate. The inset shows the central part of the unweighted invariant mass distribution }
\label{fig:mgg}
\end{figure}

The diphoton invariant mass is reconstructed from the energies measured by the calorimeter and the position of the primary vertex. In ATLAS, the primary vertex of the hard interaction is identified exploiting the directions of flight of the photons as determined with the longitudinal segmentation of the electromagnetic calorimeter. The CMS calorimeter has no pointing information and the vertex is identified from the kinematic properties of the tracks associated with that vertex and their correlation with the diphoton kinematics. 

In CMS, a multivariate regression algorithm is used to extract the photon energy and a photon-by-photon estimate of the uncertainty in that measurement. This information is used in a boosted decision tree (BDT) together with photon and vertex quality variables and kinematic variables. The  BDT is trained to separate signal and background events and its output is used to assign the events in an optimal way to four categories with different signal-to-background ratio.

In ATLAS, the non-VBF events are separated in nine categories defined by the rapidity of the photons, the component of the diphoton transverse orthogonal to the axis defined by the difference between the two photon
momenta~\cite{Ackerstaff:1997rc, Vesterinen:2008hx} and  the presence or not of converted photons.

The background in each category is estimated from data by fitting the measured diphoton mass spectrum with a background model with free shape and normalization. This background model is chosen by requiring that the potential residual bias is smaller than 20\% of the statistical accuracy of the fit. The statistical analysis of the data is done with an unbinned likelihood function in each category.

The distribution of the invariant mass of the diphoton candidates summed an all categories is shown in Fig.~\ref{fig:mgg}. In order to exploit the different sensitivity of each category, the events are weighted with category-dependent factors reflecting the different signal-to-background ratios. In ATLAS, the weights are defined  as ln(1 + S/B) where S is 90\% of the expected signal and B is the integral of the background fit in the window containing S. In CMS, the weights proportional to S/(S + B), where S and B are defined [almost] as in ATLAS and the weights are normalized such that the integral of the weighted signal model matches the number of signal events given by the best fit. An excess near 125 GeV appears clearly in both the weighted and unweighted distributions.

\subsection{Statistical analysis}
A common statistical procedure~\cite{LHC-HCG-Report} for the interpretation of the SM Higgs boson searches 
has been developed by ATLAS and CMS. Data and background predictions are compared
to the expected SM Higgs boson signal and the ratio $\mu$ between the measured and predicted signal 
strength is evaluated. The background-only hypothesis corresponds to $\mu=0$ while a positive
value of $\mu$ significantly different from zero indicates the presence of a signal. The signal expected for a 
SM Higgs boson is $\mu=1$.  At each mass, possible values of $\mu$ are tested with a test statistics based on the profile
 likelihood ratio~\cite{Cowan:2010st} to extracts the information on the signal strength from a full likelihood 
 fit to the data. The likelihood function includes all the parameters describing the data and all the parameters that 
 describe the systematic uncertainties and their correlations.

\begin{figure}[h!]
\begin{center}
\begin{tabular} {c}
\includegraphics[width=0.80\textwidth]{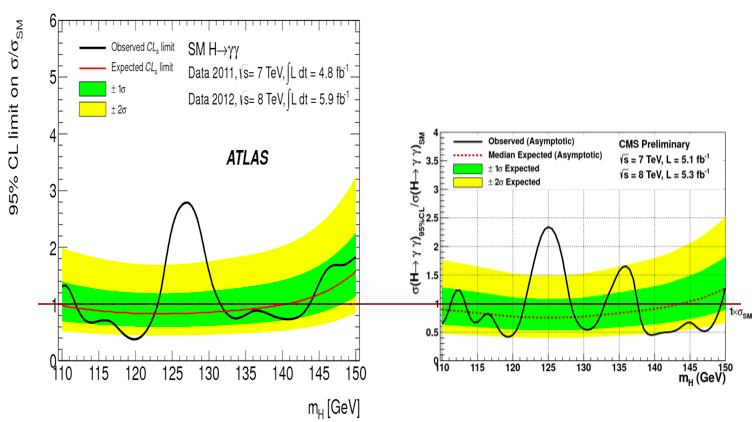}\\
\includegraphics[width=0.80\textwidth]{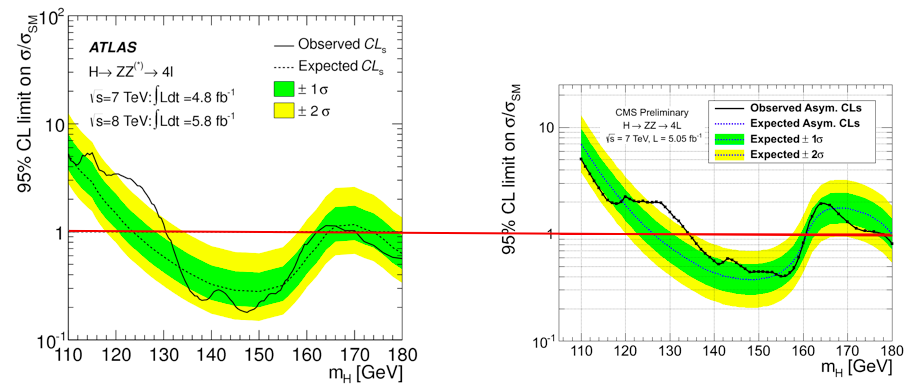}
\end{tabular}
\end{center}
\caption{ Top: $\gamma\gamma$ channel. Bottom: $4\ell$ channel. Left: ATLAS. Right: CMS. The expected exclusion limit 
$\mu_{95}$ computed in the background only hypothesis is shown together with its 1$\sigma$ (green) and 2$\sigma$ (yellow) bands as a function of the Higgs boson mass. The black line shows the observed $\mu_{95}$.}
\label{fig:muexclusion}
\end{figure}

The interpretation strategy is based on the modified frequentist 
criterion~\cite{Junk:1999kv,Read:2002hq} CLs. A value of $\mu$ is excluded at 95\% CL when CLs is less than 5\%.
For each mass, the value of $\mu$ excluded at 95\% CL, $\mu_{95}$, 
is computed. In practice a scan in steps of a fraction of the mass resolution is done for each channel.
Figure~\ref{fig:muexclusion} shows the value of $\mu_{95}$ as a function
of the Higgs boson mass for the $\gamma\gamma$ and $4\ell$ channels.

In this plot, nearby mass points are correlated with a correlation length that is given by
the mass resolution of about 1-2\%. The data follow roughly the expected value of
$\mu_{95}$ with the exception of the region at ${\rm m}_{\rm H}\simeq 125$ GeV where a peak exceeding
the 2$\sigma$ band is observed in all distributions.

\begin{figure}[h!]
\begin{tabular} {cc}
\centering\includegraphics[width=0.48\textwidth]{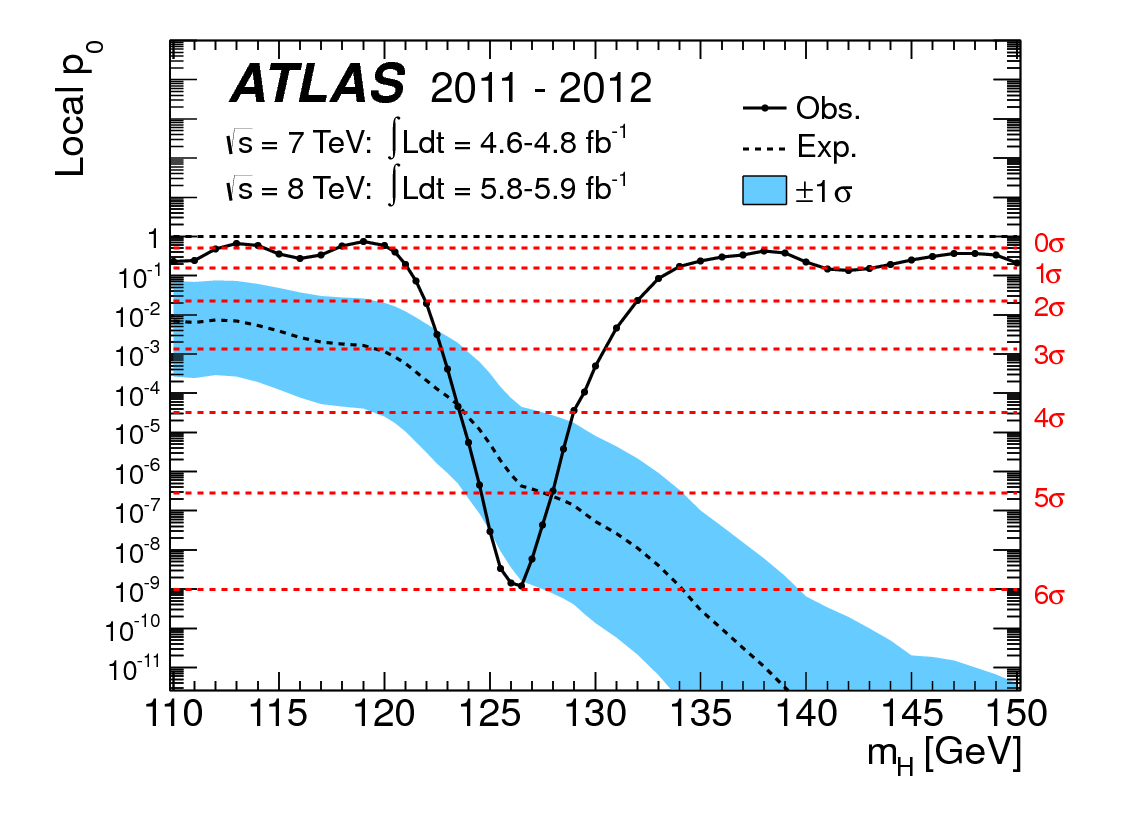}
&
\centering \includegraphics[width=0.36\textwidth]{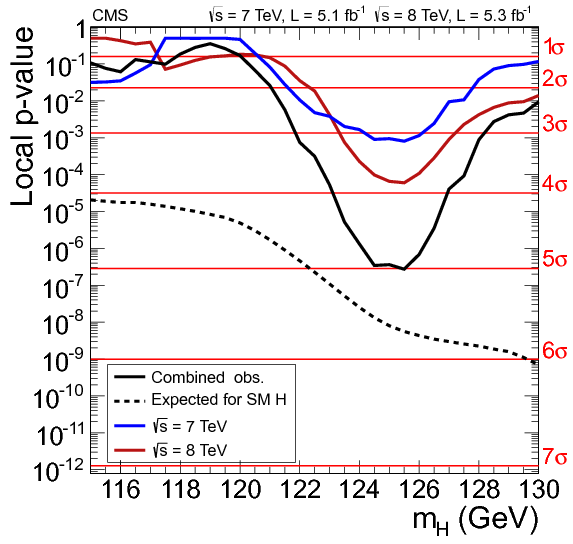}
\end{tabular}
\caption{Expected and observed local p-value as function of the Higgs boson mass for the combination of the channels of each experiment. Left: ATLAS, the band shows the  $\pm 1\sigma$ statistical fluctuation of the expected p-value. Right: CMS the observed p-values of the 7 TeV and 8 TeV data sets are also shown. }
\label{fig:pvalues}
\end{figure}

The significance of the excess is quantified by 
the probability for a background fluctuation to  be at least as large as the observed excess. 
This  local p-value is shown in Fig~\ref{fig:pvalues} for the combination of channels
presented by ATLAS and CMS. The combination assumes the SM branching fractions. 

At ${\rm m}_{\rm H}\simeq125$ GeV ATLAS expect $5\sigma$ and observe $6\sigma$
while CMS expect $6\sigma$ and observe $5\sigma$. A clear signal is established in each experiment. The combination 
of the two experiments has a global 
significance in large excess of the $5\sigma$ value that is the value conventionally required for claiming an observation.
Since this new resonance decays to two photons, it must be a boson with spin different from 1~\cite{Landau,Yang}.
Its mass can be measured by fitting the signal strength as a function of the mass in the most sensitive channels:  $\gamma\gamma$ and $4\ell$. This fit allows the signal strength in each channel to float independently in order to reduce the model dependence. As a result, CMS quotes a mass of $125.3\pm0.4\pm0.5$ GeV and ATLAS quotes a mass of $126.0\pm0.4\pm0.4$ GeV. In both experiments, the main systematic error comes from the energy scales of electrons and photons evaluated from a comparison of data and simulation at the Z peak.

\begin{figure}[h!]
\begin{tabular} {ccc}
 \includegraphics[width=0.28\textwidth]{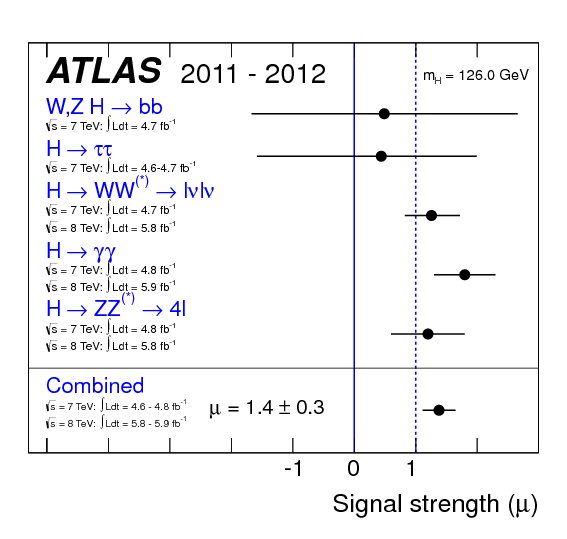}
&
  \includegraphics[width=0.25\textwidth]{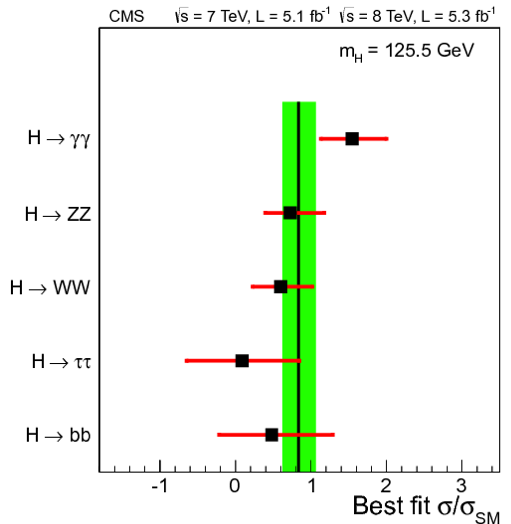}
&
\includegraphics[width=0.28\textwidth]{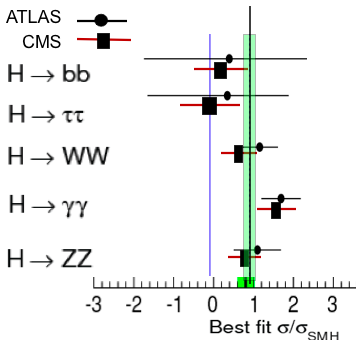}

\\
 (a)&(b)&(c)
\end{tabular}
\caption{Measurements of the signal strength parameter $\mu$ for  for the individual channels.The horizontal bars indicate the $\pm 1 \sigma$ uncertainties on $\mu$ for individual modes; they include both statistical and systematic uncertainties. (a): ATLAS assume ${\rm m}_{\rm H}=126$ GeV. (b): CMS assume ${\rm m}_{\rm H}=125.5$ GeV, the vertical band shows the overall $\mu$ value from a global fit assuming the SM branching fractions. (c) Comparison of the ATLAS and CMS measurements shown in (a) and (b). }
\label{fig:muchannels}
\end{figure}

The best-fit signal strength is fit for each search channel independently at the measured value of the mass of the boson. The results of these fits are shown in Fig~\ref{fig:muchannels}. There is large consistency between the measurements of the two experiments. The bb and $\tau\tau$ channels, with low sensitivity, are compatible with $\mu=0$ and with $\mu=1$. The WW and ZZ channels have strengths close to $\mu=1$ in both experiments, with ATLAS consistently larger than CMS. The strength of the $\gamma\gamma$ channel is larger than 1 in both experiments:  ATLAS measures $\mu=1.8\pm0.5$ and CMS $\mu=1.6\pm0.4$ . The average of the two values gives $\mu=1.7\pm0.3$, some two sigmas above the SM expectation. 

\begin{figure}[h!]
\begin{tabular} {cc}
\centering\includegraphics[width=0.48\textwidth]{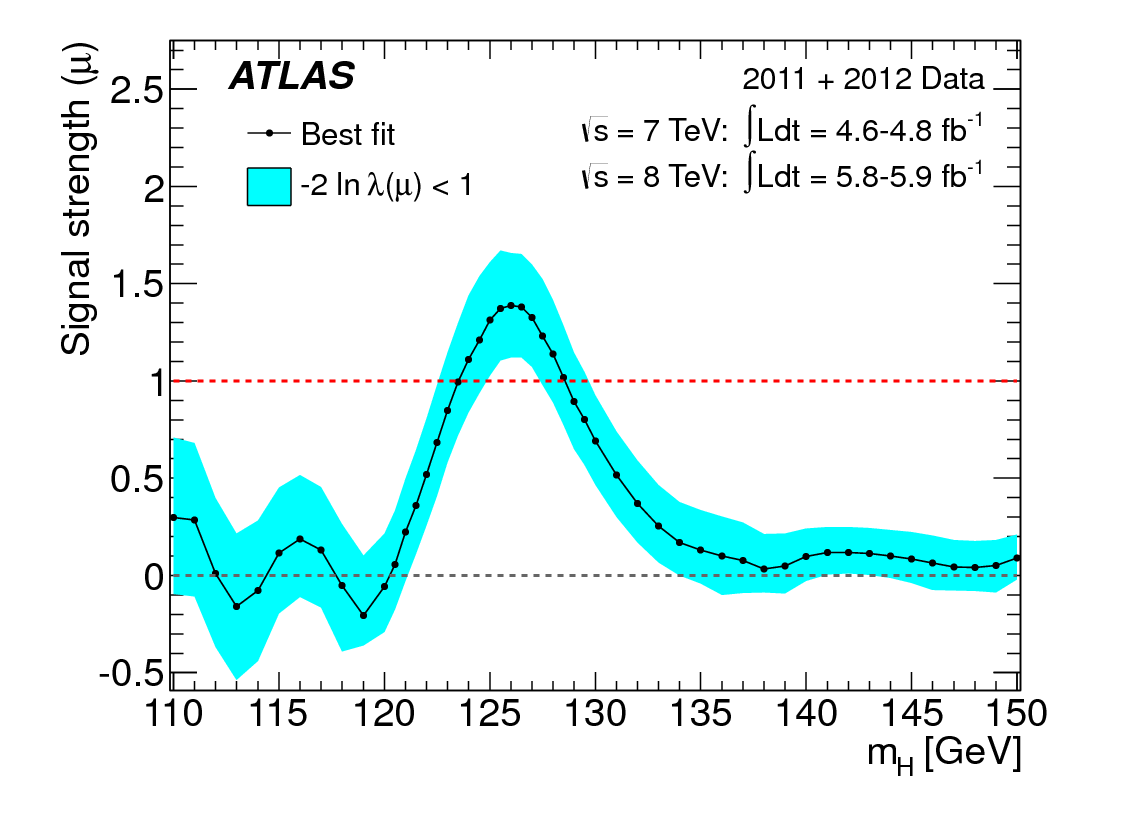}
&
\centering \includegraphics[width=0.38\textwidth]{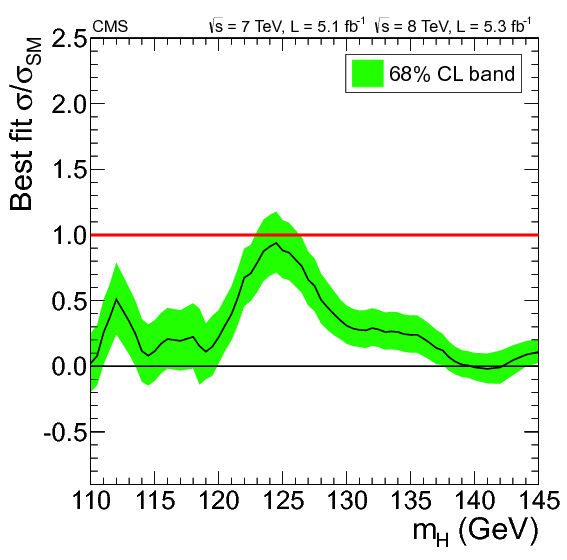}
\end{tabular}
\caption{Best-fit signal strength as a function of the Higgs boson mass hypothesis for the full combination of the 2011 and 2012 data. The band shows the  $\pm 1 \sigma$ uncertainty. Left: ATLAS. Right: CMS. }
\label{fig:mu}
\end{figure}

An important quantity is the best fit value of  the signal strength $\mu$ as a function of the Higgs boson mass for the combination of all search channels in each experiment. The combination assumes the SM branching fractions. This quantity is shown in Fig.~\ref{fig:mu}. The signal strength is compatible with the SM expectation:  ATLAS measures a slight excess $\mu=1.4\pm0.3$ and 
CMS  measures $\mu=0.87\pm0.23$. The average gives $\mu=1.1\pm0.2$.

\section{Conclusions}
The Standard Model has passed the first scrutiny by LHC, which probed for the first time the TeV scale with data collected in 2011. The Higgs particle was the main missing block of the SM. The new boson found by ATLAS and CMS in the range of masses preferred by the precision electroweak tests is a spectacular confirmation of the SM framework.  Still the SM leaves too many open questions to be considered a complete description of Nature.

Supersymmetry,  considered as one of the most natural extensions of the SM, has been tested already with 2011 data. Direct searches exclude constrained SUSY models. The room for natural supersymmetry is quite restricted but scenarios with light stops or sbottoms are still open. 

We look forward to the analysis of the new data collected by LHC in 2012 and to the higher energy run that will start after the energy upgrade to 13.5 TeV in 2015.

\bibliographystyle{utcaps}
\bibliography{pa-bib}
\end{document}